# Ultrafast energy relaxation of quantum dot-generated 2D hot electrons

Dongsung T. Park[1], Dongkun Kim[2], Uhjin Kim [2], Hwanchul Jung [3], Juho Choi[1], Cheolhee Han[1], Yunchul Chung[3], H.-S. Sim[1], V. Umansky[4], Hyoungsoon Choi [1*], Hyung Kook Choi [2*]

[1] *Department of Physics, KAIST, Daejeon 34141, Republic of Korea*

[2] *Department of Physics, Research Institute of Physics and Chemistry, Jeonbuk National University, Jeonju 54896, Republic of Korea*

[3] *Department of Physics, Pusan National University, Busan 46241, Republic of Korea*

[4] *Department of Condensed Matter Physics, Weizmann Institute of Science, Rehovot 76100, Israel*

[*] h.choi@kaist.ac.kr

[*] hkchoi@jbnu.ac.kr

**Through a series of transverse magnetic focusing experiments, we show that hot electrons in a two-dimensional electron gas system undergo an ultrafast relaxation when generated by a quantum dot (QD) instead of a quantum point contact (QPC). We find here that QPC hot electrons were well described by the non-interacting Fermi gas model for excitations up to $1.5$ meV above the Fermi level of $7.44$ meV, whereas QD hot electrons exhibited an energy loss quadratic to the excitation. The energy relaxation was a sizeable fraction of the tested excitations, up to about $55\%$. With the proposal that the hot electrons are relaxed by the QD immediately after emission, we present a toy model in which a capacitive coupling between the QD and its leads results in a finite, ultrafast energy relaxation.**

Quasiparticles form the basis upon which a vast array of many-body systems, ranging from band metals to topological insulators, are understood in condensed matter physics [1–3]. When extending beyond the ground state, the properties of quasiparticle excitations become crucial in describing system behaviors. In particular, the single electron excited above a Fermi sea has become the archetype of fermionic excitations. These excitations are commonly referred to as hot electrons and have been continuously studied over the past few decades due to their role in understanding coherent quantum devices [4–17]. Specifically, their relaxation mechanisms are a central yet controversial topic, still remaining to be fully explained [18–23].

In mesoscopic physics, single hot electrons have been realized using the discrete energy levels of a quantum dot (QD) [24]. Naturally, QDs have found widespread use as energy filters, as both single hot electron sources and energy spectrometers [6,13–17,25,26]. In the context of energy relaxation, these applications presume that QD-generated hot electrons have sufficiently long lifetimes. However, this presumption has recently been challenged by reports of ultrafast relaxation of QD hot electrons in integer quantum Hall systems [25,26]. Such relaxations had previously been veiled, possibly by the conventional use of quantum point contacts (QPC) in generating the hot electrons [4–7,10–12,16,17,27]. This posits the question of how QPC and QD hot electrons differ, and whether the latter have non-vanishing lifetimes.

Here, we report an ultrafast relaxation of QD hot electrons in the low magnetic field regime. A series of transverse magnetic focusing (TMF) experiments was performed to measure the energy of hot electrons after propagating through an open two-dimensional reservoir. After realizing a TMF device with a QD electron source, the experiments were repeated for hot electrons of varying energy levels. The results revealed an ultrafast relaxation at length scales much shorter than the conventional ballistic length. Such relaxations were absent in the reference QPC-sourced experiments performed with the same device. We suggest that the relaxation occurs near the QD and present a toy model capable of relaxing the hot electrons through a capacitive coupling between the QD and the leads.

**Experimental Setup.** The device shown in Fig. 1(a) was fabricated on a GaAs/AlGaAs heterojunction containing a two-dimensional electron gas (2DEG) 75 nm below the surface with density $n = 2.08 \times 10^{11}$ cm$^{-2}$ and mobility $\mu = 3.8 \times 10^6$ cm$^2$/Vs. Metallic Schottky gates of 75 nm width were deposited on the surface using standard electron beam lithography. A QD was formed by depleting the 2DEG underneath four neighboring gates, and the plunger gate voltage ($V_{PG}$) was used to modulate the QD energy level. The QPCs were defined using three gates rather than the traditional two. A trench gate screens the electric field of the split gates and sharpens the confinement potential [28,29], which widens the inter-subband energy separation and allows the QPC to retain its conductance quantization for a wider range of bias voltage. An AC excitation of $v_{ac} \leq 10$ μV$_{rms}$ at 987.6 Hz and a DC voltage bias ($V_{DC}$) were summed through a bias-tee and supplied to the relevant source reservoirs (Fig. 1(a), red).

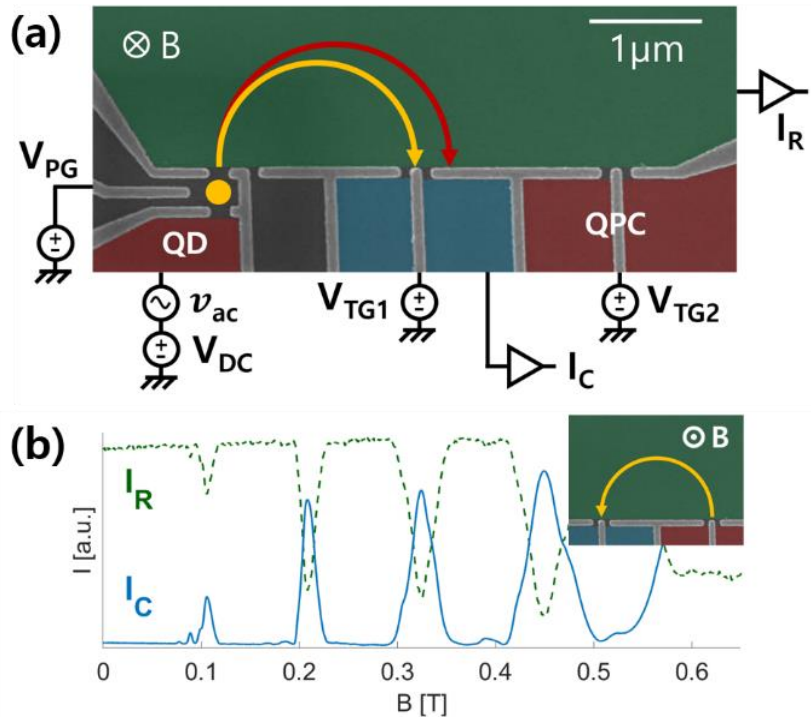

**Figure 1. Device image and QPC TMF.** (a) False-colored scanning electron micrograph of the device. A QD and two trench-gated QPCs were defined on a GaAs/AlGaAs heterostructure. The QD or the right QPC was used as the emitter (red reservoirs) and the center QPC as the collector (blue reservoir). The measurement scheme is shown for a QD-emitter TMF experiment. Focused electrons (yellow line) are drained past the collector ($I_C$), while the rest are drained through the open reservoir ($I_R$). Electrons are focused further away at higher energies (red line). (b) In a typical QPC TMF spectrum, peaks in $I_C$ (solid blue) and dips in $I_R$ (dashed green) appear as multiples of $B_0$. The right QPC was used for the QPC TMF experiment (inset).

Electrons travelling through the open reservoir emitted by the QD or QPC source (Fig. 1(a), green) either pass through the collector reservoir (Fig. 1(a), blue) or are deflected back into the open reservoir. The currents drained at the collector ($I_C$) and open reservoir ($I_R$) were measured simultaneously using two lock-in amplifiers, each fit with a homemade transimpedance preamplifier [30]. Two types of devices were made with different distances between the QD and QPCs: 1 μm and 1.5 μm. Figures 1 and 2 feature a 1.5 μm device, tested at an electron temperature of ≈250 mK; figures 3 and 4, a 1 μm device at ≈100 mK. The presented results were reproducible in other devices as well.

**QPC Transverse Magnetic Focusing.** Transverse magnetic focusing can act as an inherent spectrometer by taking advantage of the momentum dependence of the cyclotron radius [10,27,31]. A charged particle feels a Lorentz force while moving under a magnetic field.

This force deflects the free particle into a circular orbit that has a radius dependent on the particle's momentum and the magnetic field, i.e. the cyclotron radius. Consider a free electron emitted with momentum $p$ in the direction orthogonal to a collector located at distance $L$. The electron is collected when a perpendicular magnetic field of strength $B_0$ is present (Fig. 1(a), yellow curve):

$$B_0 = \frac{p}{eL/2}, \qquad (1)$$

where $e$ is the absolute charge of the electron. If there is a reflecting barrier between the emitter and the collector, the electron skips along the barrier and is collected at multiples of $B_0$. When a collimated beam of electrons is emitted, the collected current plotted against the magnetic field is called the focusing spectrum of the beam. Figure 1(b) is a typical example, measured from our sample using the QPC emitter (Fig. 1(b), inset). The roughly periodic appearance of peaks corresponds to the multiples of some $B_0$, which we call the focusing field.

At the same field, however, electrons with greater kinetic energy have a larger cyclotron radius and will not be collected (Fig. 1(a), red curve). In the massive Fermi gas model, an electron only has kinetic energy, i.e. $E \propto p^2$. Therefore, hot electrons with energy $-eV_{DC}$ above the Fermi level $E_F$ will have a modified focusing field $B_0(V_{DC})$ related to that of equilibrium electrons by

$$\frac{B_0(V_{DC})}{B_0} = \sqrt{1 - \frac{eV_{DC}}{E_F}}, \qquad (2)$$

where we have assumed the effective mass to be constant [10]. By the above principle, TMF can be used as an energy spectrometer, akin to the optical monochromator [10,27]. Energy loss in hot electrons can then be observed as a deviation from Eq. (2).

**QD Transverse Magnetic Focusing.** We replaced the conventional QPC emitter with a QD to test if QD hot electrons retain their energies during emission and propagation in the 2D reservoir. However, performing QD TMF is nontrivial due to the effect of the magnetic field on the QD [24]. A magnetic field shifts the QD energy levels and rearranges the order of the orbitals. Consequently, the valence electron's energy becomes unpredictable, as seen in the non-monotonic shifts in QD conductance by the magnetic field in Fig. 2(a). Naïve use of a QD emitter leads to undesirable changes in the hot electron energy on the order of the QD excitation energy.

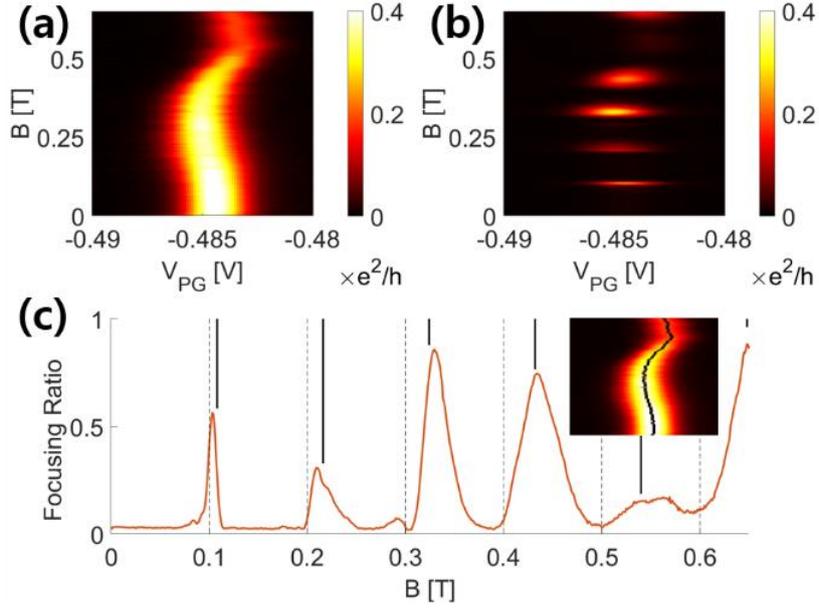

**Figure 2. Quantum dot TMF.** (a) QD conductance $(I_C + I_R)/v_{ac}$ shows the Coulomb blockade peak shifting erratically with the magnetic field. The magnetic field also changes the valence electron coupling to the leads. (b) Focusing conductance $I_C/v_{ac}$ is the product of the focusing spectrum and QD conductance. (c) $I_C/(I_R + I_C)$ traced along the Coulomb blockade peak (inset) gives us the focusing ratio (orange) normalized for changes in the QD–lead coupling strength. Geometric uncertainties in the QD and QPC positions result in a slight mismatch between the observed $B_0$ (solid lines) and the expected values (dashed lines).

Here, the magnetic energy shifts were cancelled by continually adjusting $V_{PG}$ to align the transmitting QD level with the Fermi level of the leads. The Fermi level of a 2DEG does not change in low magnetic fields and thus serves as an appropriate energy reference. At the appropriate $V_{PG}$, the QD level aligns with the Fermi levels, and QD conductance is maximized. We retrieved the focusing spectrum for a fixed energy level by examining $I_C$, which is approximately the product of the focusing spectrum and the QD conductance (Fig. 2(b)), at $V_{PG}$ values where $I_C + I_R$ exhibited Coulomb blockade peaks (Fig. 2(c), inset). Figure 2(c) is the resulting spectrum, normalized for changes in QD coupling strength by plotting $I_C/(I_C + I_R)$ against the magnetic field. This normalized focusing spectrum, which we call the focusing ratio, was used for the remainder of our analysis.

The QD focusing spectrum was similar to that from the QPC TMF. The first peak of the focusing spectrum was well predicted by Eq. (1) with the expected lineshape [31], but the skipping orbit peaks exhibited several deviations. First, the focusing field slightly deviated from the predicted value; the dominant source of error in this case can be attributed to geometric uncertainty. The lithographic gap in the QD and QPC in our experiment could cause a deviation in $B_0$ up to $\approx 10$ %. Fortunately, this systematic error was unimportant since Eq. (2) only deals with relative changes in $B_0$. The peculiar lineshape of the later peaks can be attributed to interference between multiple paths [31], boundary specularity and roughness issues in skipping orbits [32], and the increasing relevance of quantum Hall edge states [33]. Therefore, we restricted our interest to the first peak in order to avoid such irrelevant effects.

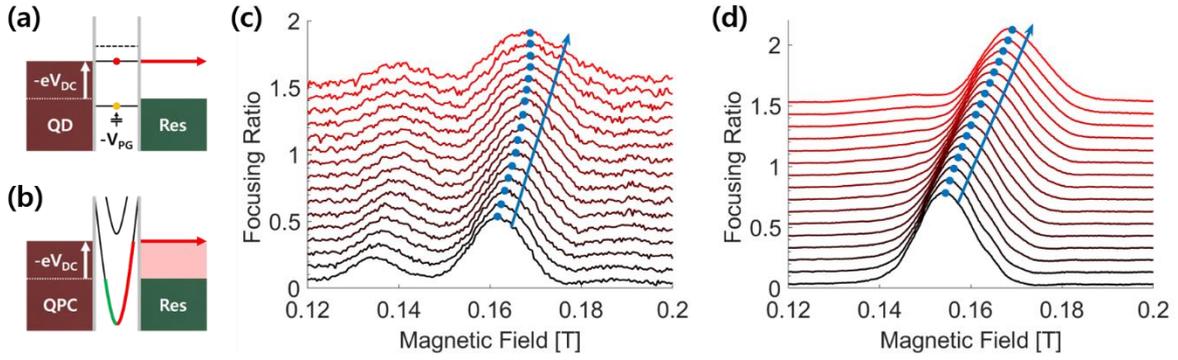

**Figure 3. Hot electron TMF.** (a) $V_{PG}$ aligns the QD level with the electrochemical potential $E_F - eV_{DC}$ of the emitter reservoir. The QD was tuned to tunnel only through a single level (solid line with red circle). (b) For an appropriate $V_{TG}$, the QPC channeled electrons only through its lowest subband for all relevant $V_{DC}$ values. The minimum of the transmitting subband was always lower than the grounded 2DEG electrochemical potential, and the mimimum of the next subband was higher than the biased level. (c) QD TMF spectra for varying levels of $V_{DC}$. Each redder line corresponds to a spectrum with an additional bias of $-0.1$ mV, drawn with an offset increment of $+0.1$. The focusing peak extracted by a fit (blue circles) tracks how $B_0(V_{DC})$ changes. At low $V_{DC}$, the peak shifts rather linearly. At higher $V_{DC}$, however, the peaks do not line up with the low $V_{DC}$ trend (blue arrow). (d) QPC TMF spectra for varying levels of $V_{DC}$ drawn similar to (b). The focusing peak shifts almost linearly.

**Hot Electron Transverse Magnetic Focusing.** A clear difference between QD TMF and QPC TMF was observed in hot electron experiments. The energies of QD hot electrons were controlled by aligning the QD level to the biased electrochemical potential $E_f - eV_{DC}$ (Fig. 3(a), Supplementary Fig. 1). The excitation levels were tuned to be greater than the estimated value of the thermal energy, so that only the main QD energy level would tunnel electrons. The resulting focusing spectra for $V_{DC}$ from 0 mV to $-1.5$ mV are shown in Fig. 3(c), with redder lines corresponding to hotter electrons. At small $V_{DC}$, the focusing peak shifted linearly with $V_{DC}$, while at larger $V_{DC}$, the peak shift exhibited a visible nonlinearity (Fig. 3(c)). As a reference to traditional results, QPC TMF was also performed on the same device with care taken to maintain only one channel between the biased and grounded electrochemical potentials (Fig. 3(b), Supplementary Fig. 2). This condition was satisfied by setting $V_{TG}$ to where the conductance plateau was maintained for all relevant bias levels, i.e. $g = 2e^2/h$. In stark contrast to the QD results, the QPC TMF peak shifted almost linearly (Fig. 3(d), Supplementary Figs. 3–5).

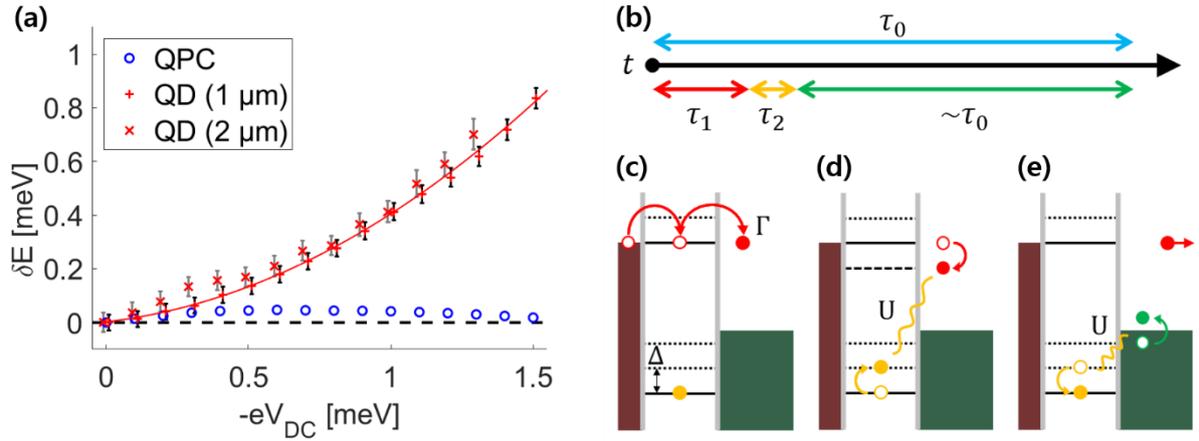

**Figure 4. Energy relaxation and toy model.** (a) Energy loss $\delta E$ during transit from the QD to the collector, calculated from the relative focusing peak shift and Eq. (3). The 2 μm QD TMF data was measured using the previous emitter QPC as the collector; the 2 μm and 1 μm dots are horizontally offset by 0.01 mV for clarity. Error bars were assigned by summing the Coulomb peak width and the focusing peak position fitting error range by full width at half maximum; error bars for the QPC data are smaller than the marker. Unlike the QPC TMF, both QD TMF cases exhibit a similar, increasing energy loss sizeable to $-eV_{DC}$. A quadratic fit to the 1 μm data is provided: $\delta E_{fit} = 0.28(-eV_{DC})^2 + 0.12(-eV_{DC})$. (b) An electron is tunneled every $\tau_0$. (c) During the first $\tau_1$, an electron tunnels through the QD. (d) For a short time $\tau_2$, the emitted electron excites the QD through a capacitive interaction. (e) In the remainder of $\tau_0 - (\tau_1 + \tau_2) \sim \tau_0$, (e) the QD relaxes with the leads to reach its ground state again.

**Energy Relaxation.** Analysis of the focusing peak shift revealed a strong energy relaxation of hot electrons only present in QD TMF. The kinetic energy of hot electrons can be calculated using Eq. (2). The $-eV_{DC}$ term is the hot electron energy within the QD, and the focusing field gives us the hot electron energy in the open reservoir. The difference in these energies provides the energy deficit $\delta E(V_{DC})$:

$$\frac{\delta E(V_{DC})}{E_F} = \left(\frac{B_0(V_{DC})}{B_0}\right)^2 - \left(1 - \frac{eV_{DC}}{E_F}\right). \tag{3}$$

Very little energy deficit was present in the QPC TMF spectra (Fig. 4(a), blue circles), which agrees with past reports [10,13,34]. However, a sizeable portion of the hot electron energy was lost in the QD TMF case (Fig. 4(a), red crosses). Moreover, the energy deficit in QD TMF was insensitive to doubling the focusing distance from 1 μm to 2 μm. Similar results were obtained for skipping orbit peaks as well.

From this discrepancy, we can speculate two possible scenarios: either QPC electrons relax less, or QD electrons relax more. Specifically, the first scenario implies that QPC-generated hot electrons are suppressed from relaxation in the 2D reservoir. This is unlikely to be the discriminant, since a relaxation mechanism in the 2DEG would lead to greater relaxations for longer focusing distances. In contrast, our results show that the QD hot electrons still retain a sizeable portion of their initial energy, as if the relaxation had abruptly stopped before depleting the electrons of their excess energy. Therefore, the second scenario is more likely, in which QD-generated hot electrons experience stronger relaxations. In particular, we expect the hot electrons to lose energy near the QD at a length scale much shorter than 1 µm.

**Toy Model.** The above phenomena can be qualitatively captured by the following toy model. We extend the usual QD model to incorporate a capacitive interaction between the leads and the QD. An electron tunnels through the QD with period $\tau_0 = e/I \gtrsim 260$ ps (Fig. 4(b), blue). The electron dwells inside the QD for a duration on the order of the QD level lifetime $\tau_1 = \hbar/\Gamma \approx 65$ ps (Fig. 4(b), red), where $\Gamma$ is the QD–leads tunneling strength (Fig. 4(c)). Right after tunneling, the hot electron may capacitively interact with the QD within some interaction length scale $l_{rel}$. This length scale can be larger than the bulk 2D screening length $\approx$ 10 nm but must be much shorter than the focusing length $\approx$ 1.5 µm—the local screening length may be larger than the bulk 2D value due to gating effects, such as 2DEG depletion and lowered electron density. Such interaction occurs within an ultrafast timescale $\tau_2 = l_{rel}/v_f \ll 10$ ps (Fig. 4(b), orange), where $v_f$ is the Fermi velocity. During this time, the hot electron can transfer part of its energy to the QD (Fig. 4(d)) before propagating away. The excited QD is then left to relax with the leads (Fig. 4(e)) for $\tau_0 \gg \tau_1, \tau_2$ before returning to its ground state to repeat the process (Fig. 4(b), green). The expected hot electron energy loss is given by the expected QD excitation during $\tau_2$. Predictions from a semi-classical rate equation for a reasonable set of parameter values qualitatively resemble the observed energy loss (Supplementary Note 1 and Fig. 6).

After a careful comparison between QD and QPC emitters in magnetic focusing experiments, we conclude that an ultrafast relaxation exists for QD-generated 2D hot electrons on the order of the excitation. Such a relaxation was absent in QPC TMF, as reported in previous studies. Although a more detailed model is required for better quantitative explanation, we found that a simple toy model can present a similar relaxation through a QD-mediated capacitive interaction. In particular, our relaxation model does not invoke the presence of a magnetic field and is symmetric under time reversal, suggesting that a QD may not only relax the hot electrons leaving it, but also those entering it. This relaxation may have gone unnoticed, since the energy loss becomes increasingly pronounced with higher excitations.

Our results suggest that QDs may not be reliable energy filters in 2D systems for energies larger than the QD level spacing. If a QD has limitations in creating a monoenergetic beam of electrons, then it may be possible that a QD also has similar limitations in accurately measuring the energies of lead electrons. In future research, a better non-QD spectrometer will be necessary in order to eliminate the inherent limitations of TMF; the peak broadening from beam collimation heavily burdens the task of tracking the precise hot electron energy. Nevertheless, we believe that our experiment confirms the presence of a large and ultrafast energy relaxation in QD hot electrons, an observation that will be important to further studies on quasiparticle relaxations, especially near local potential traps or impurities.

## References

[1]  P. Wölfle, Reports Prog. Phys. **81**, 032501 (2018).

[2]  A. Sommerfeld, Zeitschrift Für Phys. **47**, 1 (1928).

[3]  R. B. Laughlin, Phys. Rev. Lett. **50**, 1395 (1983).

[4]  U. Sivan, M. Heiblum, and C. P. Umbach, Phys. Rev. Lett. **63**, 992 (1989).

[5]  B. Laikhtman, U. Sivan, A. Yacoby, C. P. Umbach, M. Heiblum, J. A. Kash, and H. Shtrikman, Phys. Rev. Lett. **65**, 2181 (1990).

[6]  C. Altimiras, H. Le Sueur, U. Gennser, A. Cavanna, D. Mailly, and F. Pierre, Phys. Rev. Lett. **105**, 226804 (2010).

[7]  T. Otsuka, Y. Sugihara, J. Yoneda, T. Nakajima, and S. Tarucha, J. Phys. Soc. Japan **83**, 014710 (2014).

[8]  A. Marguerite, C. Cabart, C. Wahl, B. Roussel, V. Freulon, D. Ferraro, C. Grenier, J. M. Berroir, B. Plaçais, T. Jonckheere, J. Rech, T. Martin, P. Degiovanni, A. Cavanna, Y. Jin, and G. Fève, Phys. Rev. B **94**, 115311 (2016).

[9]  C. Bäuerle, D. Christian Glattli, T. Meunier, F. Portier, P. Roche, P. Roulleau, S. Takada,

and X. Waintal, Reports Prog. Phys. **81**, 056503 (2018).

[10] J. G. Williamson, H. Van Houten, C. W. J. Beenakker, M. E. I. Broekaart, L. I. A. Spendeler, B. J. Van Wees, and C. T. Foxon, Phys. Rev. B **41**, 1207 (1990).

[11] R. I. Hornsey, J. R. A. Cleaver, and H. Ahmed, Phys. Rev. B **48**, 14679 (1993).

[12] F. Müller, B. Lengeler, T. Schäpers, J. Appenzeller, A. Förster, T. Klocke, and H. Lüth, Phys. Rev. B **51**, 5099 (1995).

[13] T. Kobayashi, S. Sasaki, T. Fujisawa, Y. Tokura, and T. Akazaki, Phys. Status Solidi C **5**, 162 (2008).

[14] F. Hohls, M. Pepper, J. P. Griffiths, G. A. C. Jones, and D. A. Ritchie, Appl. Phys. Lett. **89**, 212103 (2006).

[15] C. Rössler, S. Burkhard, T. Krähenmann, M. Röösli, P. Märki, J. Basset, T. Ihn, K. Ensslin, C. Reichl, and W. Wegscheider, Phys. Rev. B - Condens. Matter Mater. Phys. **90**, 081302(R) (2014).

[16] C. Altimiras, H. Le Sueur, U. Gennser, A. Cavanna, D. Mailly, and F. Pierre, Nat. Phys. **6**, 34 (2009).

[17] H. Le Sueur, C. Altimiras, U. Gennser, A. Cavanna, D. Mailly, and F. Pierre, Phys. Rev. Lett. **105**, 056803 (2010).

[18] G. F. Giuliani and J. J. Quinn, Phys. Rev. B **26**, 4421 (1982).

[19] I. P. Levkivskyi and E. V. Sukhorukov, Phys. Rev. B - Condens. Matter Mater. Phys. **85**, 075309 (2012).

[20] T. Karzig, A. Levchenko, L. I. Glazman, and F. Von Oppen, New J. Phys. **14**, 105009 (2012).

[21] D. Ferraro, B. Roussel, C. Cabart, E. Thibierge, G. Fève, C. Grenier, and P. Degiovanni, Phys. Rev. Lett. **113**, 166403 (2014).

[22] J. Lee, J. E. Han, S. Xiao, J. Song, J. L. Reno, and J. P. Bird, Nat. Nanotechnol. **9**, 101 (2014).

[23] Q. Weng, S. Komiyama, L. Yang, Z. An, P. Chen, S. A. Biehs, Y. Kajihara, and W. Lu, Science (80-. ). **360**, 775 (2018).

[24]  L. P. Kouwenhoven, D. G. Austing, and S. Tarucha, Reports Prog. Phys. **64**, 701 (2001).

[25]  R. H. Rodriguez, F. D. Parmentier, P. Roulleau, U. Gennser, A. Cavanna, F. Portier, D. Mailly, and P. Roche, ArXiv E-Prints arXiv:1903.05919 (2019).

[26]  T. Krähenmann, S. G. Fischer, M. Röösli, T. Ihn, C. Reichl, W. Wegscheider, K. Ensslin, Y. Gefen, and Y. Meir, Nat. Commun. **10**, 3915 (2019).

[27]  J. G. Williamson, H. van Houten, C. W. J. Beenakker, M. E. I. Broekaart, L. I. A. Spendeler, B. J. van Wees, and C. T. Foxon, Surf. Sci. **229**, 303 (1990).

[28]  Y. J. Um, Y. H. Oh, M. Seo, S. Lee, Y. Chung, N. Kim, V. Umansky, and D. Mahalu, Appl. Phys. Lett. **100**, 183502 (2012).

[29]  M. Seo, C. Hong, S. Y. Lee, H. K. Choi, N. Kim, Y. Chung, V. Umansky, and D. Mahalu, Sci. Rep. **4**, 3806 (2014).

[30]  A. V. Kretinin and Y. Chung, Rev. Sci. Instrum. **83**, 084704 (2012).

[31]  H. Van Houten, C. W. J. Beenakker, J. G. Williamson, M. E. I. Broekaart, P. H. M. Van Loosdrecht, B. J. Van Wees, J. E. Mooij, C. T. Foxon, and J. J. Harris, Phys. Rev. B **39**, 8556 (1989).

[32]  K. Tsukagoshi, S. Takaoka, K. Murase, K. Gamo, and S. Namba, Appl. Phys. Lett. **62**, 1609 (1993).

[33]  T. Stegmann, D. E. Wolf, and A. Lorke, New J. Phys. **15**, 113047 (2013).

[34]  T. M. Chen, M. Pepper, I. Farrer, D. A. Ritchie, and G. A. C. Jones, Appl. Phys. Lett. **103**, 093503 (2013).

**Supplementary Materials**

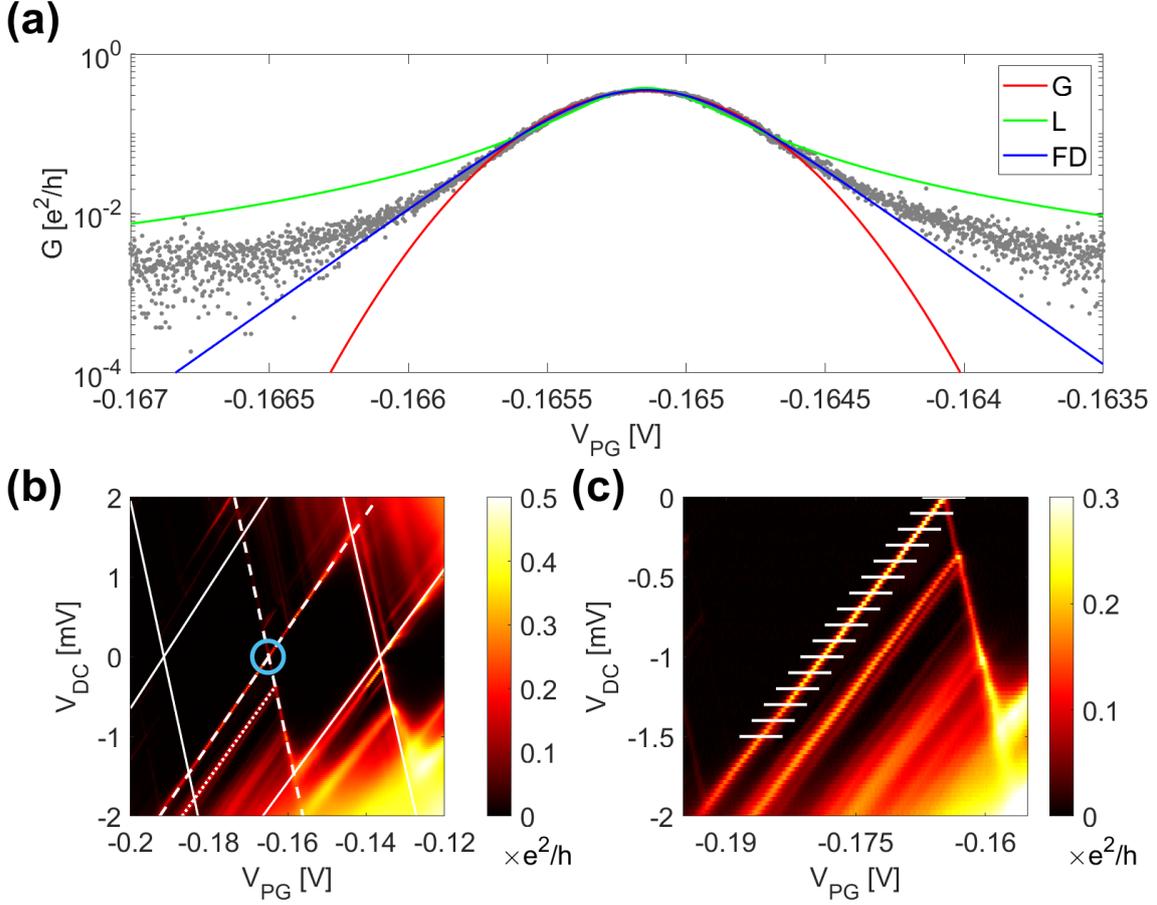

**Supplementary Figure 1. QD Characterization.** [1] (a) Coulomb blockade peak for the $n^{th}$ level from the QD in Figs. 3 and 4 of the main manuscript. The raw data (gray dots) are fit with Gaussian (red, G), Lorentzian (green, L), and the thermal (blue, FD) curves of form y ∝ sech$^2(x)$. We see that the closest fit is the thermal curve. The full width at half maximum (FWHM) of the peak is $\Delta V_{PG} = 0.624$ mV. The QD coupling was tuned as symmetrically as possible. (b) Coulomb diamond for the $n^{th}$ QD energy level (dashed cross) used in Figs. 3 and 4 of the main manuscript. At the main level (blue circle), the QD level $E$ shifts by $\alpha := -\partial E/\partial V_{PG} = 54.8$ meV/V$_{PG}$, which gives $\alpha \Delta V_{PG} = 34.2$ μeV. The $n-1^{th}$ level (left, solid cross) is visible at lower conductance ranges. The charging energy is $E_C = 1.47$ meV, and the first excitation (bottom, dotted line) is $E_X = 0.37$ meV above the ground state. (c) The edge of the Coulomb diamond with a positive slope on the $V_{DC} - V_{PG}$ axis corresponds to the case in which the QD level follows the electrochemical potential of the biased source. The QD hot electron TMF experiments were repeated along the white stiches on the Coulomb diamond edge.

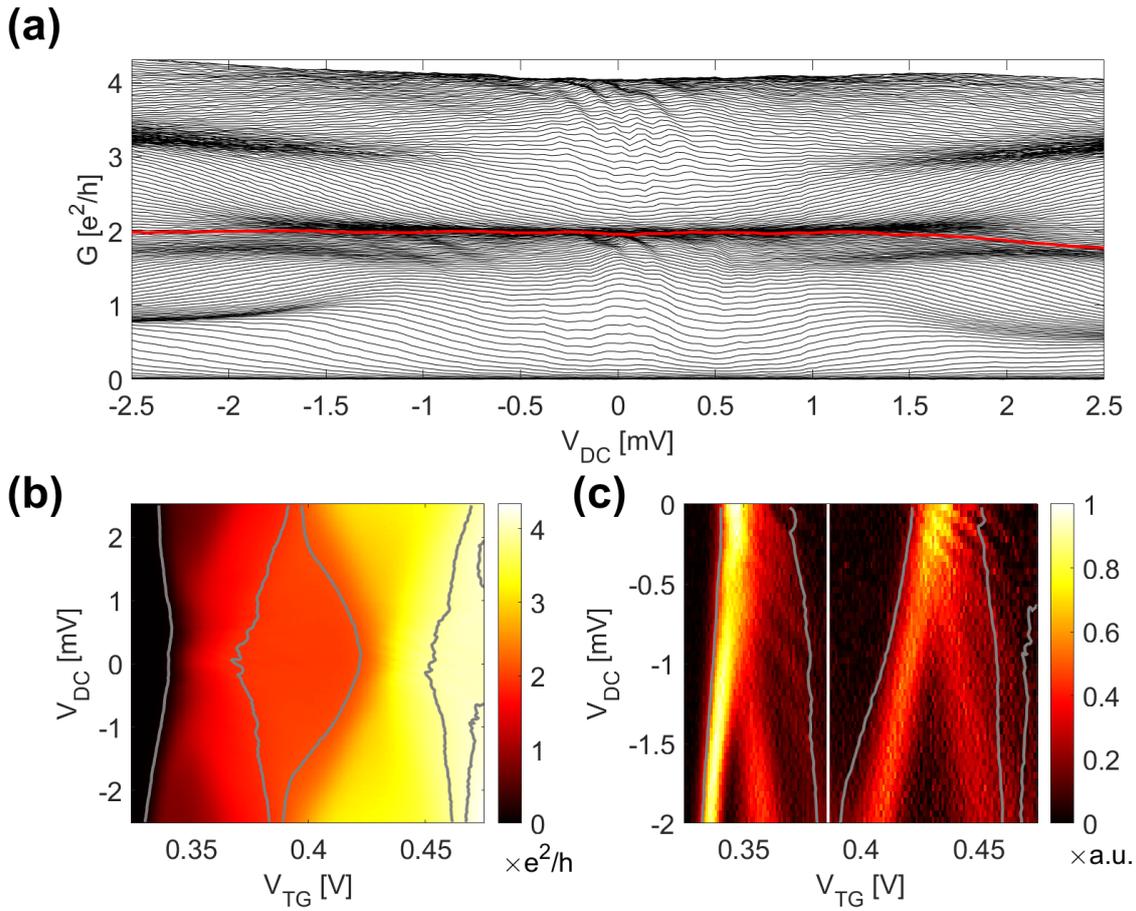

**Supplementary Figure 2. QPC Characterization.** [2] (a) Conductance of the QPC from Figs. 3 and 4 of the main manuscript drawn as a function of $V_{DC}$ in constant increments of $V_{TG}$. $V_{TG} = +0.384$ mV (red line) was used in the experiments as the value for which the conductance varies least when $V_{DC} < 0$. (b) Within the area in which the conductance deviates little from the quantized values $2n \times e^2/h$ (contours for $(2n \pm 0.1) \times e^2/h$ drawn in grey), the QPC is an integer-mode, bidirectional transport channel. The appropriate $V_{TG}$ may be selected by staying within the area and minimizing (c) the QPC's sensitivity $|\partial G/\partial V_{PG}|$. $V_{TG} = +0.384$ mV satisfies both conditions (white line).

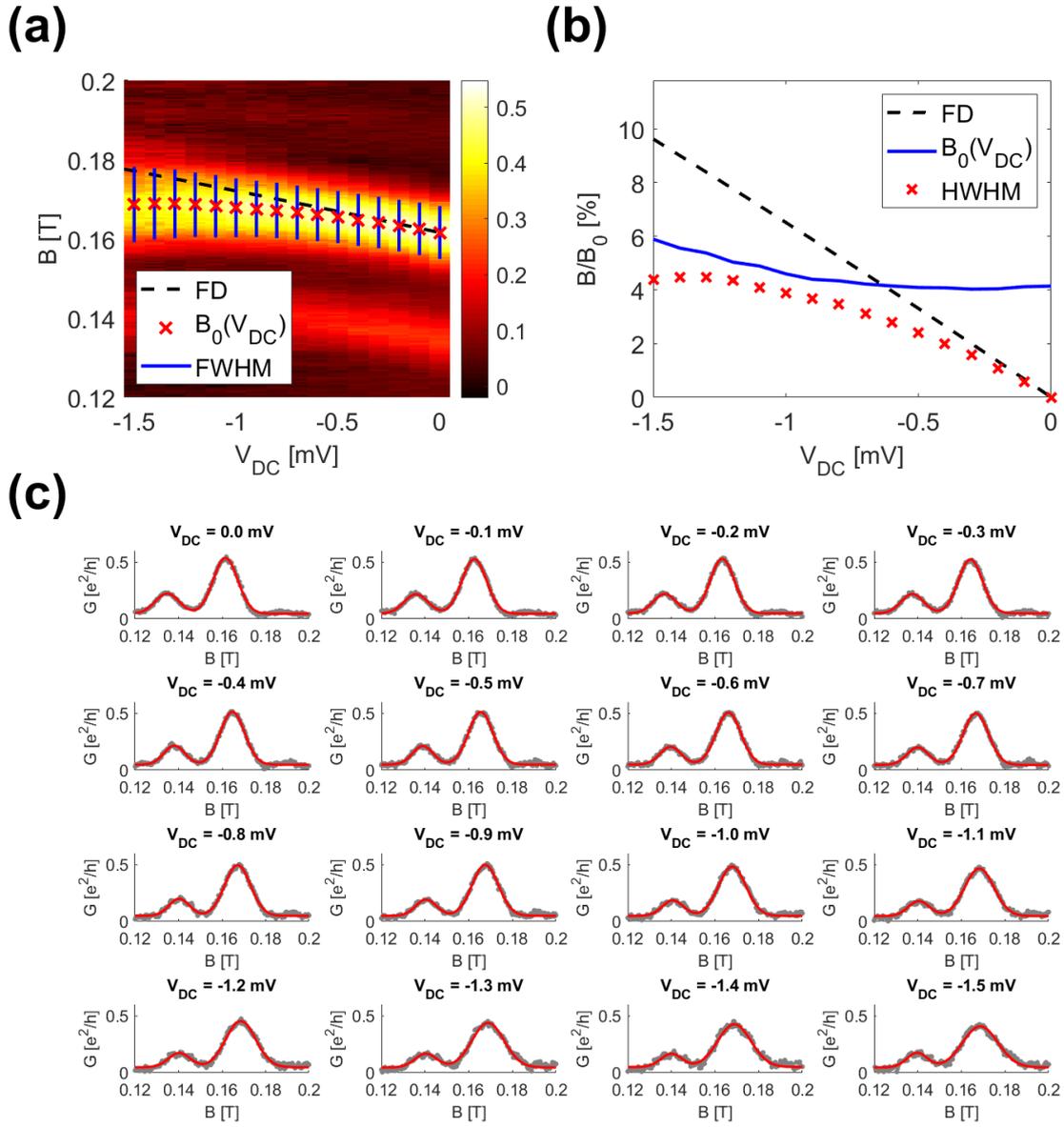

**Supplementary Figure 3. Analysis of QD hot electron TMF for a focusing distance of 1 μm.** (a) Focusing ratios for QD TMF in increments of $\Delta V_{DC} = -1$ mV. At $V_{DC} = 0$ mV, $B_0 = 162$ mT, which is within 10% of the geometrically expected value of 151 mT. The focusing peak $B_0(V_{DC})$ (red crosses) and half width at half maximum (blue line) were extracted from a double Gaussian fit. The expected peak shift in a Fermi gas (dashed lines) is given by $B_0(V_{DC}) = B_0\sqrt{1 - eV_{DC}/E_F}$, where $E_F = 7.44$ meV (measured separately). The subpeak is possibly due to branching or diffraction effects, but the latter effect is more likely since the subpeak moves together with the main peak. The fit values are summarized in (b), and the fits of individual TMF spectra are shown in (c).

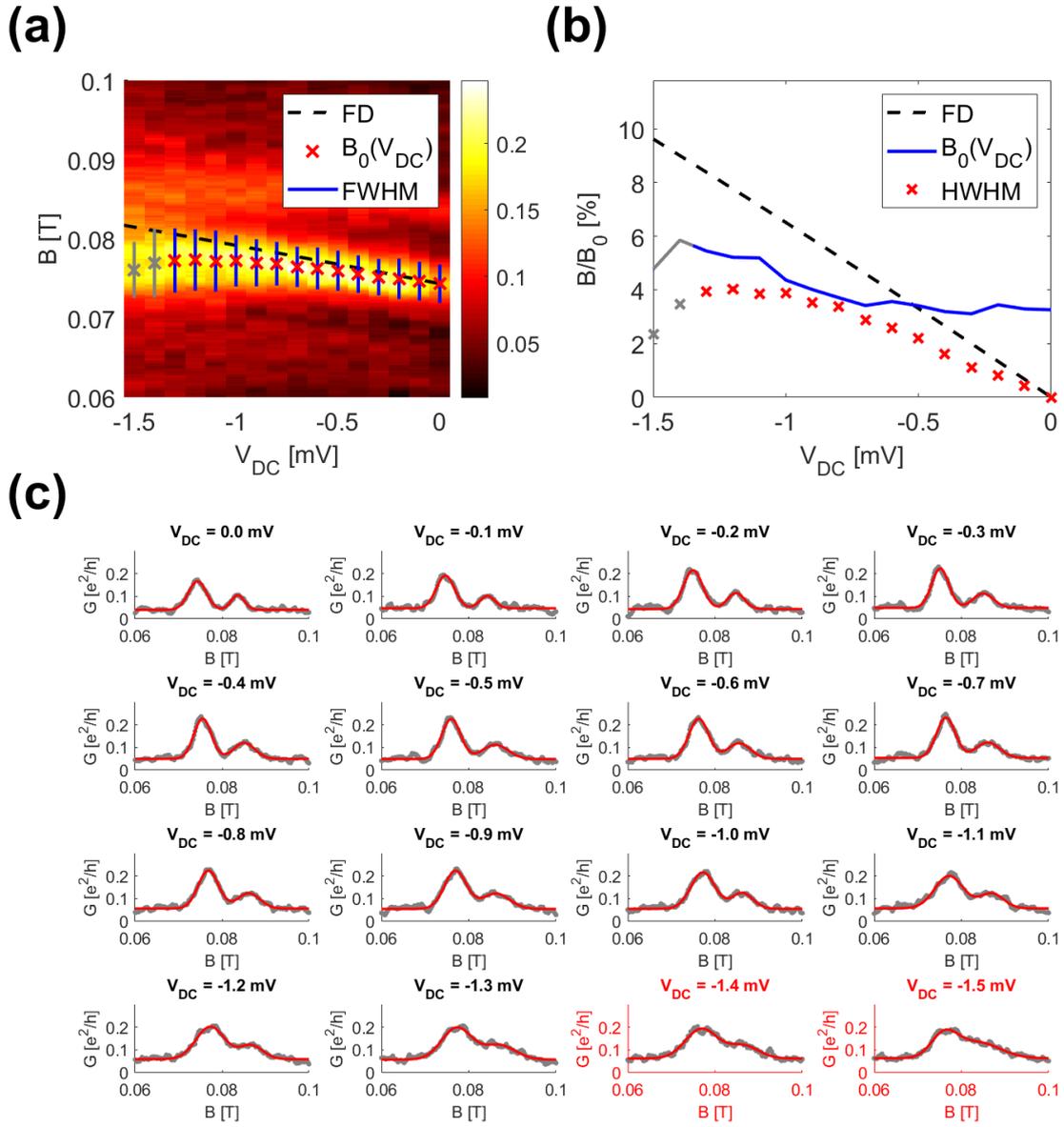

**Supplementary Figure 4. Analysis of QD hot electron TMF for a focusing distance of 2 μm.** All figures are plotted similar to Fig. S2. (a) At $V_{DC} = 0$, $B_0 = 74.3$ mT which is well within 10% of the geometrically expected value of 75.1 mT. The fit for $V_{DC} = -1.4$ mV and $-1.5$ mV are unreliable, as can be seen from (c); the two Gaussian curves are not resolved, and no minimum develops between the two peaks. Accordingly, the fit values are grayed in (a) and (b).

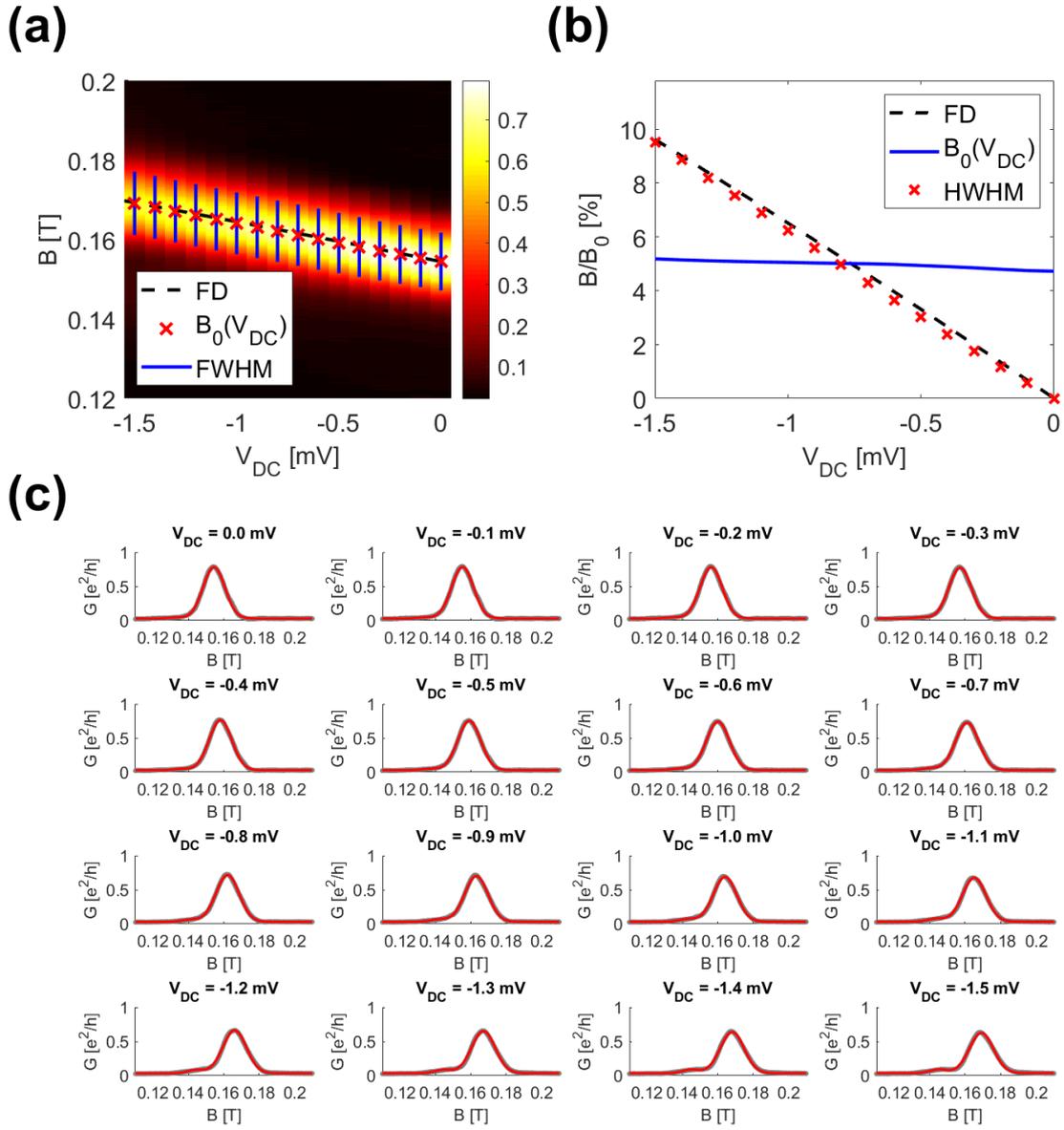

**Supplementary Figure 5. Analysis of QPC hot electron TMF for a focusing distance of 1 μm.** All figures are plotted similar to Fig. S2. (a) At $V_{DC} = 0$, $B_0 = 154$ mT which is well within 10% of the geometrically expected value of 151 mT. Although the two Gaussian curves are not resolved in most fits (c), the principal peak value is clear. We plotted the figures in (a) and (b) accordingly.

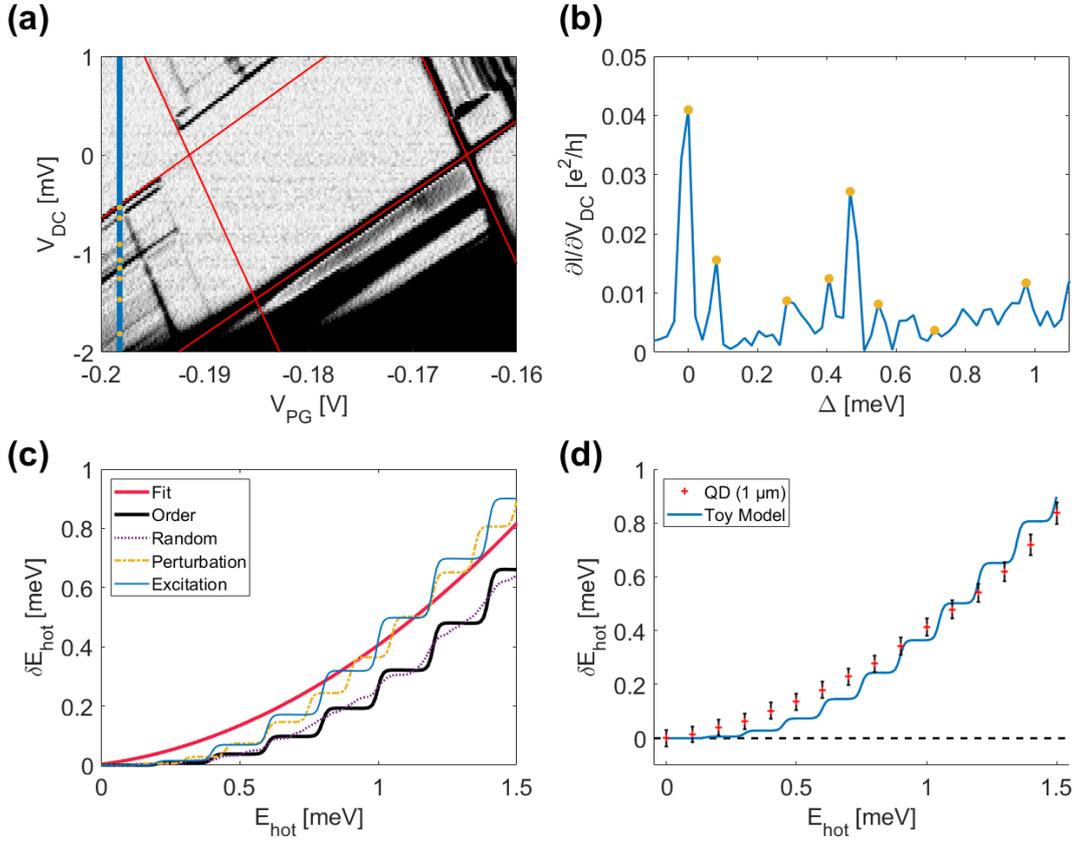

**Supplementary Figure 6. Energy loss predicted by the toy model.** (a) Coulomb diamond for the QD used in Fig. 3 and 4 of the main manuscript. QD excited states (yellow dots) were extracted at $V_{PG} = -0.1982$ mV (thick blue line). (b) QD conductance at $V_{PG} = -0.1982$ mV shown as a function of lead electrochemical potential $\mu$ from the QD ground state $E_{GS}$, i.e. $\Delta(\mu - E_{GS})$. The excitations identified from (a) are also plotted (yellow dots). The mean difference between $\Delta$ of the excited levels, and hence $\epsilon_0$, was 0.15 meV. (c) Fitted relation between the initial hot electron energy $E_{hot}$ and its energy loss during focusing. $\delta E_{hot}$ is reproduced from Fig. 4 (red). The toy model's order-of-magnitude prediction, i.e. $s = 0.042$ for $\epsilon_i = 200$ meV $\times\, i$, (solid black, Order) gives a slightly smaller relaxation than the experimental results. The stair-like feature is due to the arithmetic progression of $\epsilon_i$ in our model. Introducing a random Gaussian variation $\delta\epsilon$ with a standard deviation of $\sqrt{\langle\delta\epsilon^2\rangle} = \epsilon_0/4$ to each level, i.e. $\epsilon_i = i\epsilon_0 + \delta\epsilon_i$, quickly smoothens the line by breaking the QD excitation degeneracies (dotted purple, Random). Doubling the total perturbation, i.e. $s = 0.084$, (dashed yellow, Perturbation) raises $\delta E_{hot}$ as expected. Also, decreasing $\epsilon_0$ (solid blue, Excitation) from 200 meV to 150 meV from (b) raises $\delta E_{hot}$ as well. This is expected due to the increased degeneracies in QD excitations. (d) The blue excitation line from (c) is plotted with the 1 μm QD experimental data. We see that our toy model is capable of producing an energy relaxation with qualities similar to that of the experiment within a reasonable set of parameters. Detailed complications regarding the QD level distribution and a more accurate description of the interaction may lead to better fitting models.

**Supplementary Note 1: Toy Model**

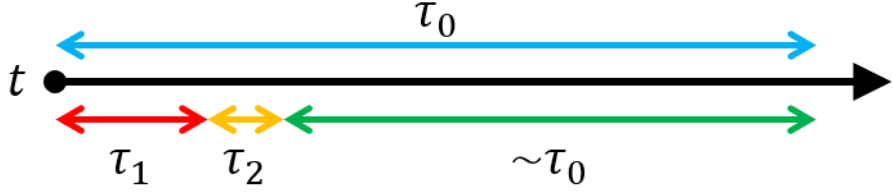

**Fig. 4(b)** from the main article.

In a current $I$, an electron is tunneled with an approximate period $\tau_0 = e/I$, where $e$ is the electron's absolute charge. For a single QD level at $E_0$ coupled to leads $\lambda = L, R$ with coupling strengths $\Gamma_{L,R}$, the equilibrium current is given by

$$I_{eq} = \frac{e}{h}\int dE \frac{\Gamma_L \Gamma_R}{(E-E_0)^2 + (\Gamma/2)^2}[f_L(E) - f_R(E)]$$

where $\Gamma = \Gamma_L + \Gamma_R$, $f_{L,R}(E) = f(E - \mu_{L,R})$, and $f(x) = (1 + \exp(x/k_B T))^{-1}$ for the thermal energy $k_B T$. The maximum current $\overline{I_{eq}}$ through a QD with symmetric coupling ($\Gamma_L = \Gamma_R$) is then simply $\overline{I_{eq}} = e/h \times \pi\Gamma/2$ in the linear response regime. From $\Gamma = \hbar\tau_1^{-1}$, where $\tau_1$ is the electron's dwell timescale in the QD, we get $4\tau_1 \leq \tau_0$. The QD is empty for the duration of $(\tau_0 - \tau_1) \sim \tau_0$. From the Coulomb peak size shown in Fig. S1(a), we know the QD is in the regime of $\Gamma \approx kT \equiv 100\ mK$ — $\Gamma \approx \alpha\Delta V_{PG}/4 \approx 10\ \mu eV$, consistent with our electron temperature. This gives us a conservative estimate of $\tau_0 \gtrsim 260$ ps. Now, consider an electron being focused; it travels $\pi L/2 \approx 1.6\ \mu m$ with a velocity of $v \gtrsim v_f = 2 \times 10^5\ ms^{-1}$, giving us a flight time of 10 ps which bounds the relaxation time $\tau_2 \ll 10$ ps. The clear separation of time scales motivates us to model the relaxation as a process involving a single hot electron.

We can imagine a process in which an electron in the lead, including the 2DEG reservoir and the hot electron, capacitively interacts with the QD for a short period of time. During the interaction, the electron may lose its energy by exciting the QD. The toy model Hamiltonian $H$ is set up as follows:

$$H = H_{dot} + H_{res} + H_{tun} + H_{int} \qquad \text{S1 (a)}$$

$$H_{dot} = \sum_i \epsilon_i n_i + \epsilon_c \sum_{i<j} n_i n_j \qquad \text{S1 (b)}$$

$$H_{res} = \sum_{k,\lambda} \epsilon_{k,\lambda} c^\dagger_{k,\lambda} c_{k,\lambda} \qquad \text{S1 (c)}$$

$$H_{tun} = \sum_{i,\lambda} t_{i,\lambda} d^\dagger_i \psi_\lambda(x=0) + h.c. \qquad \text{S1 (d)}$$

$$H_{int} = \overline{U}\sum_{i,j,\lambda} d^\dagger_j d_i \rho_\lambda(x=0). \qquad \text{S1 (e)}$$

The first three terms in Eq. S1(a), $H_0 = H_{dot} + H_{res} + H_{tun}$, give the usual QD model Hamiltonian; $H_{dot}$ counts the energy of occupied orbitals $\Sigma_i \epsilon_i n_i$ and the electrostatic repulsion energy between occupied orbitals $\epsilon_c \Sigma_{i<j} n_i n_j$ where $\epsilon_i$ is the energy of the $i^{th}$ QD orbital, $n_i = d^\dagger_i d_i$ for $d^\dagger_i$ being the creation operator of said level, and $\epsilon_c$ the capacitive contribution to the

QD energy; $H_{res}$ counts the energy of the lead reservoir $\Sigma_{k,\lambda} \epsilon_{k,\lambda} c^\dagger_{k,\lambda} c_{k,\lambda}$ where $\epsilon_{k,\lambda}$ is the energy of an electron with momentum $\hbar k$ in lead $\lambda$ and $c^\dagger_{k,\lambda}$ is the creation operator of said electron; and $H_{tun}$ tunnels lead $\lambda$ electrons near the QD ($x = 0$) into orbital $i$ with hopping strength $t_{i,\lambda}$ via $\Sigma_{i,\lambda} d^\dagger_i \psi_\lambda(x=0)$ where $\psi^\dagger_\lambda(x=0) = \frac{1}{L}\Sigma_k c^\dagger_{k,\lambda}$ is the creation operator of lead $\lambda$ near the QD—the Hermitian conjugate $h.c.$ provides the reverse process. $H_0$ is well-understood and explains the tunneling behaviors between the QD and the leads [3].

The novel part here is the additional term $H_{int}$ that can excite an electron from orbital $i$ to orbital $j$ through a capacitive interaction $\bar{U}$ with a charge density of lead $\lambda$ electrons near the QD $\rho_\lambda(x=0) = \psi^\dagger_\lambda(x=0)\psi_\lambda(x=0)$, i.e. $\bar{U} \Sigma_{i,j,\lambda} d^\dagger_j d_i \rho_\lambda(x=0)$. Expanding out Eq. 1(e) in the momentum basis and writing $U = \bar{U}/L^2$, we get

$$H_{int} = U \sum_{i,j,k,k',\lambda} d^\dagger_j d_i c^\dagger_{k',\lambda} c_{k,\lambda}.$$

By allowing lead electrons to exchange energies with the QD, our toy model relaxes hot electrons by exciting a nearby QD. The interaction is short range $l_{hot} < \pi L/2$ as implied in $\rho_\lambda(x=0)$. Consider a state $|\sigma, \boldsymbol{q}\rangle$ in which a QD is in some configuration $\sigma$ alongside a hot electron with momentum $\hbar \boldsymbol{q}$. Through $H_{int}$, an electron in the occupied QD orbital $i$ can hop to an unoccupied orbital $j$ by changing the momentum of the hot electron to $\hbar \boldsymbol{q}'$. The system with a new QD configuration $\sigma'$ can be written as $|\sigma', \boldsymbol{q}'\rangle = d^\dagger_j d_i c^\dagger_{q'} c_q |\sigma, \boldsymbol{q}\rangle$. Using Fermi's golden rule, we may calculate the transition rate $\gamma_{hot}$ from QD state $\sigma$ to state $\sigma'$ for all possible $\boldsymbol{q}$ and $\boldsymbol{q}'$:

$$\gamma^{\sigma' \leftarrow \sigma}_{hot} = \frac{2\pi}{\hbar} \Sigma_{q',q} |\langle \sigma', \boldsymbol{q}' | H_{int} | \sigma, \boldsymbol{q} \rangle|^2 W_{\sigma,q} \delta(E_{\sigma,q} - E_{\sigma',q'}), \qquad \text{S2}$$

where $W_{\sigma,q}$ is thermal distribution weighting for state $|\sigma, \boldsymbol{q}\rangle$ and $E_{\sigma q} = E_\sigma + \epsilon_q$ is the system energy with QD and hot electron contributions $E_\sigma$ and $\epsilon_q$, respectively. The right-hand side reduces to $\frac{2\pi}{\hbar} \rho U^2 (1 - f(\epsilon_{q'}))(1 - f(\epsilon_q))$ where $\rho = E_f^{-1}$ is the normalized 2DEG density of states. For a series of similar transitions, we may consider a reference state with energy $E = E_{\sigma,q} = E_{\sigma',q'}$ which we will define as the sum of the QD ground state energy $E_{GS}$ and the initial hot electron energy $\epsilon_k$. Then, we may rewrite Eq. S2 as

$$\gamma^{\sigma' \leftarrow \sigma}_{hot} = \frac{2\pi}{\hbar} \rho U^2 (1 - f(\epsilon_k - \Delta_{\sigma'}))(1 - f(\epsilon_k - \Delta_\sigma)), \qquad \text{S3}$$

where $\Delta_\sigma = E_\sigma - E_{GS}$ is the excitation energy for the QD state $\sigma$. Similarly, the rate $\gamma_{QD}$ at which the QD is relaxed by creating an electron–hole pair in the reservoir is

$$\gamma^{\sigma' \leftarrow \sigma}_{QD} = \frac{2\pi}{\hbar} \rho^2 U^2 |\Delta_{\sigma'} - \Delta_\sigma| \Theta(\Delta_\sigma - \Delta_{\sigma'}), \qquad \text{S4}$$

where we have approximated for $|\Delta_\sigma| \gg kT$ and $\Theta$ is the Heaviside step function. This rate is doubled if we consider the QD to be relaxing through both reservoirs.

These rates are plugged into the Master equation to solve for the expected excitation in the QD and hence the energy loss of the hot electron:

$$\frac{dP_\sigma}{dt} = \sum_{\sigma'} \left( \gamma_{hot}^{\sigma \leftarrow \sigma'} P_{\sigma'} - \gamma_{hot}^{\sigma' \leftarrow \sigma} P_\sigma \right). \qquad \text{S5}$$

The differential equation is solved for the initial condition $P_\sigma(t=0) = \delta_{\sigma,GS}$ over the time $\tau_2$, leading to the expected energy loss $\delta E$ in the hot electron by

$$\delta E = \sum_\sigma \Delta_\sigma P_\sigma(\tau_2). \qquad \text{S6}$$

In the meantime, the QD relaxing via the lead reservoir can be modeled by similar means:

$$\frac{dP_\sigma}{dt} = \sum_{\sigma'} \left( \gamma_{rel}^{\sigma \leftarrow \sigma'} P_{\sigma'} - \gamma_{rel}^{\sigma' \leftarrow \sigma} P_\sigma \right). \qquad \text{S7}$$

Since $\gamma_{hot}/\gamma_{rel} = E_f/\Delta_{ij} \gg 1$, we may approximate to the leading term by solving Eq. S5 only for the duration of $\tau_2$ and then switching over to Eq. S7 for the remainder of $\tau_0 - (\tau_1 + \tau_2) \approx \tau_0$.

Our resulting model for hot electron energy relaxation has two types of fitting parameters. The first is a dimensionless parameter $s = \gamma_{hot} \tau_2 \propto U^2 l_{hot}$, which dictates how 'long' $P_i$ evolves, and the second parameter is the energy distribution $\{\Delta_\sigma\}$ of the QD which dictates how much energy the hot electron may lose per interaction. For the QDs used in this experiment, $\epsilon_{N+1} - \epsilon_N$ is usually ~200 µeV, i.e. $\Delta_\sigma$ is approximately some multiple of ~200 µeV. A QD has a net charge of $q \lesssim e$, so the potential $U$ estimated as the Coulomb potential of an electron immediately across a depleted GaAs region of width ~150 nm is $\lesssim$ 0.8 meV; $l_{hot}$ is likely on the order of the screening length ~10 nm. This leads to an order-of-magnitude estimate that $s \sim 0.042$ and $\epsilon_0 = 200$ meV for $\epsilon_i = i \times \epsilon_0$. Calculations are shown in Fig. S6.

**References**

[1]   L. P. Kouwenhoven, D. G. Austing, and S. Tarucha, Reports Prog. Phys. **64**, 701 (2001).

[2]   C. Rössler, S. Baer, E. De Wiljes, P. L. Ardelt, T. Ihn, K. Ensslin, C. Reichl, and W. Wegscheider, New J. Phys. **13**, 113006 (2011).

[3]   H. Bruus and K. Flensberg, in *Many-Body Quantum Theory Condens. Matter Phys. An Introd.* (Oxford University Press, 2004), pp. 152–183.